\documentclass[aps,pra,floatfix,twocolumn]{revtex4}
\usepackage{amsmath}
\usepackage{bm}
\usepackage{euscript}
\usepackage{graphicx}
\graphicspath{{./Figures/}}

\def\ea0{\mbox{ $ea_0$}}

\def\mF0{\mbox{ $\mu\Phi_0$}}
\def\mF0rtHz{\mbox{ $\mu\Phi_0/\sqrt{\rm Hz}$}}
\def\rtHz{\mbox{$\sqrt{\rm Hz}$}}

\def\ecm{\mbox{ e$\cdot$cm}}

\begin{document}
\title{The prospects for an electron electric dipole moment search with ferroelectric (Eu,Ba)TiO$_3$ ceramics.}
\author{A. O. Sushkov}
\email{alex.sushkov@yale.edu} 
\affiliation{Yale University, Department of Physics, P.O. Box 208120, New Haven, CT 06520-8120}
\author{S. Eckel}
\affiliation{Yale University, Department of Physics, P.O. Box 208120, New Haven, CT 06520-8120}
\author{S. K. Lamoreaux}
\affiliation{Yale University, Department of Physics, P.O. Box 208120, New Haven, CT 06520-8120}

\begin{abstract}
We propose to use ferroelectric (Eu,Ba)TiO$_3$ ceramics just above their magnetic ordering temperature for a sensitive 
electron electric dipole moment search. We have synthesized a number of such ceramics with various Eu concentrations and measured their properties relevant for such a search: permeability, magnetization noise, and ferroelectric hysteresis loops. The results of our measurements indicate that a search for the electron electric dipole moment with Eu$_{0.5}$Ba$_{0.5}$TiO$_3$ should lead to an order of magnitude improvement on the current best limit.
\end{abstract}

\date{August 25, 2009}

\maketitle

\section{Introduction}

Searching for permanent electric dipole moments (EDMs) of particles such as electron and neutron, and atoms, such as $^{199}$Hg, is a promising way of probing fundamental physics beyond the Standard Model. Existence of a non-zero EDM is a signature of violation of discrete symmetries of parity and time-reversal invariance (and therefore CP-violation). The present best experimental upper limit on electron EDM is $1.6\times 10^{-27}\ecm$, which already constrains certain supersymmetric extensions of the Standard Model~\cite{Regan2002}.
The Standard Model predicts much smaller EDMs ($10^{-42}\ecm$ for electron, for example) due to a number of cancelations~\cite{Khriplovich1997}. However a number of theories of physics beyond the Standard Model (Supersymmetry, Grand Unification, Multi-Higgs, etc) predict electron EDM values within two orders of magnitude of the present limit~\cite{Bernreuther1991}.

\begin{figure}[h!]
    \includegraphics[width=\columnwidth]{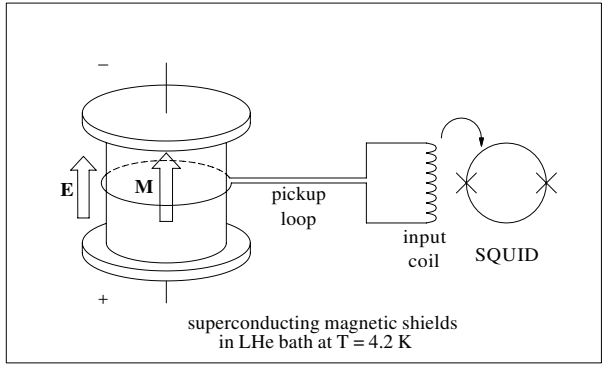}
    \caption{A schematic experimental setup for an EDM experiment.}
    \label{fig:EDMSetUp}
\end{figure}
There are two common experimental schemes of searching for an EDM: magnetic-resonance experiments, and electric-field-correlated magnetization measurements~\cite{Budker2006}. We plan to use the latter method. The electron's EDM has to point along its magnetic moment (spin). When an electric field is applied to the sample, it orients permanent electric dipole moments along the field, and hence the magnetic moments are also oriented, creating a magnetization, see Fig.~\ref{fig:EDMSetUp}. A SQUID magnetometer is used to detect this magnetization, as the electric field is reversed~\cite{Sushkov2009}. The most sensitive solid-state-based EDM search to date employed an inverse effect~\cite{Heidenreich2005}.

The EDM sensitivity of a magnetization-based EDM search can be estimated from ~\cite{Sushkov2009}
\begin{equation}
\label{eq:edm1}\delta d_e = \frac{\mu_a}{\mu-1}\frac{\delta B}{E^*},
\end{equation}
where $d_e$ is the achievable EDM limit, $\mu_a$ is the magnetic moment of the atomic species involved,  $\mu$ is the magnetic permeability of the material, $\delta B$ is the magnetic field sensitivity (taking into account a suitable demagnetizing factor), and $E^*$ is the effective electric field.
We have identified the ferroelectric Eu$_x$Ba$_{1-x}$TiO$_3$, as a very promising material for an electron EDM search. A permanent electron EDM induces a permanent EDM of the Eu$^{2+}$ ion (ground state $^8$S$_{7/2}$, configuration [Xe]4f$^7$), whose magnetic moment is $\mu_a=8\mu_B$.
The reason for using a ferroelectric is the extremely large effective electric field: $E^*\approx 10$~MV/cm, in a poled ceramic (see below).
Europium is chosen because of its large atomic number (EDM enhancement factor scales roughly as $\alpha^2Z^3$), as well as its magnetic properties (due to the 4f electrons), which give rise to a permeability of $\mu-1 \approx 0.5$ at 4.2~K.
Using the SQUID magnetometers that we have in our lab, we can achieve magnetic field sensitivity of $\delta B \approx 3$~fT/\rtHz. We found that the intrinsic magnetization noise in Eu$_{0.5}$Ba$_{0.5}$TiO$_3$ is at or below this value.
With these parameters, using ceramic Eu$_{0.5}$Ba$_{0.5}$TiO$_3$, the projected EDM sensitivity with our current design after 10 days of averaging is:
\begin{equation}
\label{eq:edm2} d_e \approx 1.5\times 10^{-28}\ecm,
\end{equation}
which is a factor of 10 improvement on the present best limit. Below we describe the measurements of the relevant properties of (Eu,Ba)TiO$_3$, on which the above estimate is based.

\section{The relevant properties of Eu$_x$Ba$_{1-x}$TiO$_3$ ceramics}

\subsection{Preparation}

The ceramics were synthesized by solid-state reaction of oxide powders Eu$_2$O$_3$, TiO$_2$, and BaCO$_3$ in a furnace with a flowing atmosphere of 5\% H$_2$, 95\% Ar (which reduces the europium to its Eu$^{2+}$ oxidation state). Ceramic disks of 13~mm diameter and 2~mm thickness were pressed in a uniaxial press at 7-ton pressure, and sintered in the same atmosphere. We synthesized and studied samples of the following compositions: EuTiO$_3$, Eu$_{0.75}$Ba$_{0.25}$TiO$_3$, Eu$_{0.5}$Ba$_{0.5}$TiO$_3$, and Eu$_{0.25}$Ba$_{0.75}$TiO$_3$. Ceramic densities were above 60\%.

\subsection{Crystal structure}
\begin{figure}[h!]
    \includegraphics[width=\columnwidth]{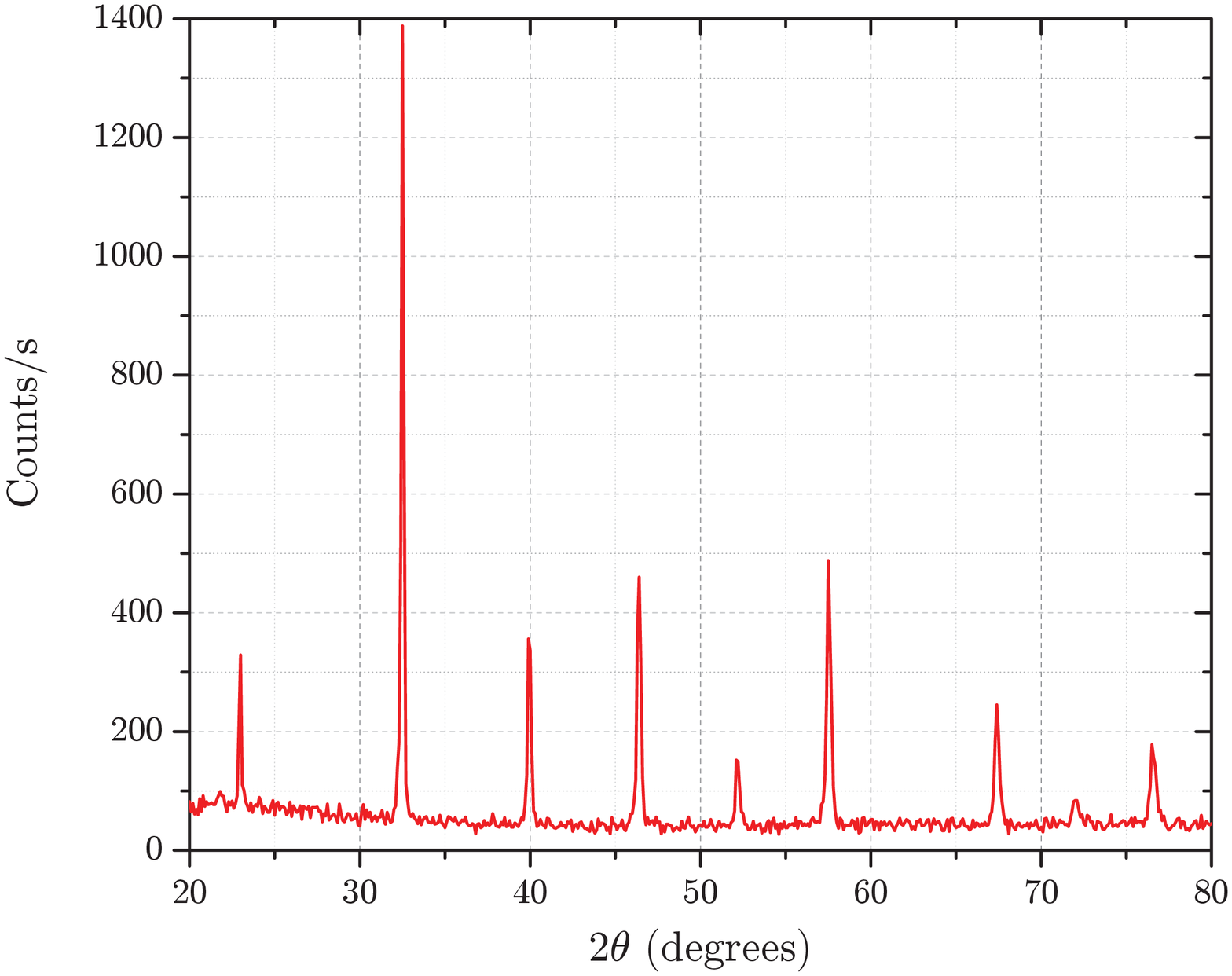}
    \caption{XRD spectrum of Eu$_{0.5}$Ba$_{0.5}$TiO$_3$ ceramic at 300~K.}
    \label{fig:XRD1}
\end{figure}
\begin{figure}[h!]
    \includegraphics[width=\columnwidth]{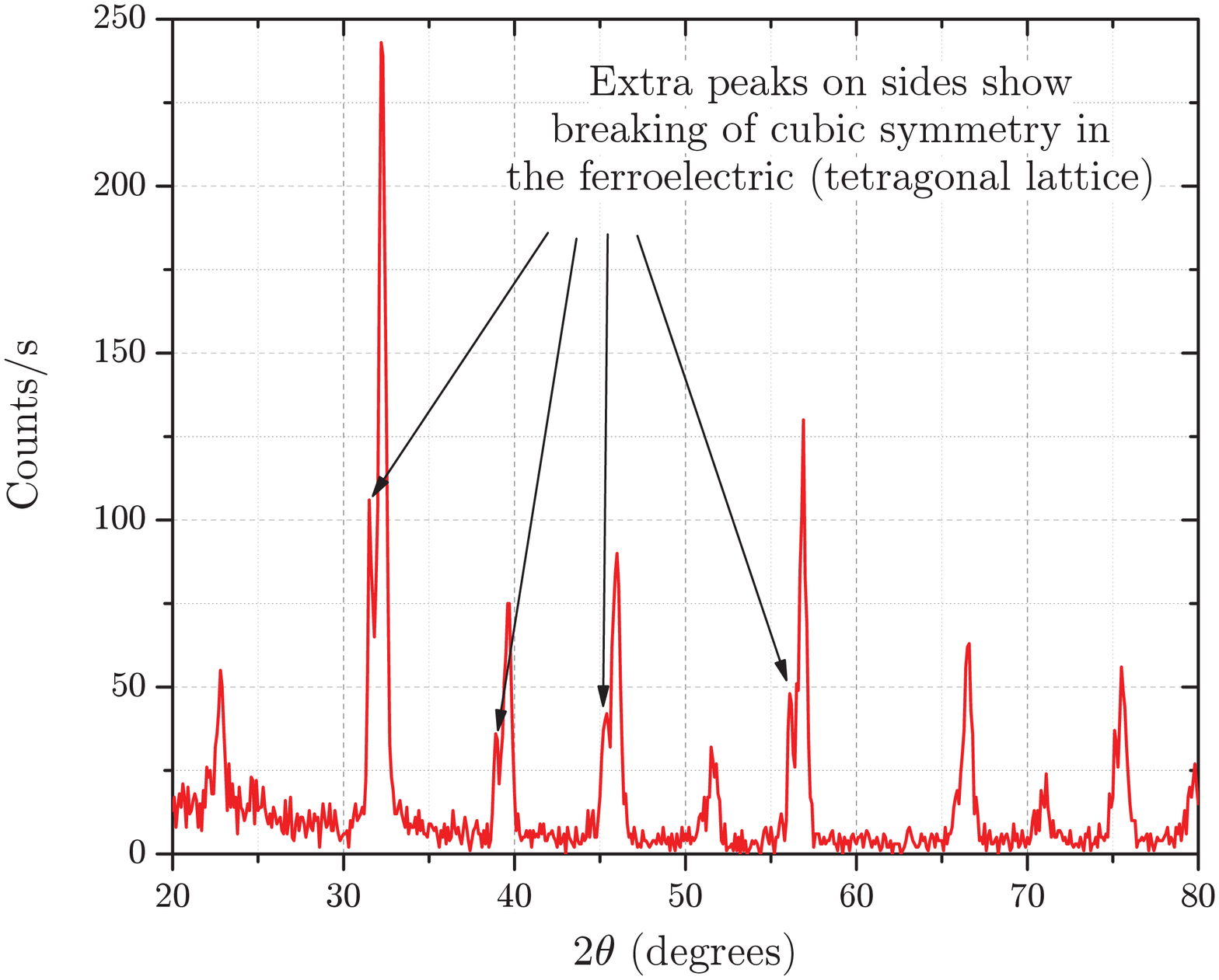}
    \caption{XRD spectrum of BaTiO$_3$ powder at 300~K.}
    \label{fig:XRD2}
\end{figure}
\begin{figure}[h!]
    \includegraphics[width=\columnwidth]{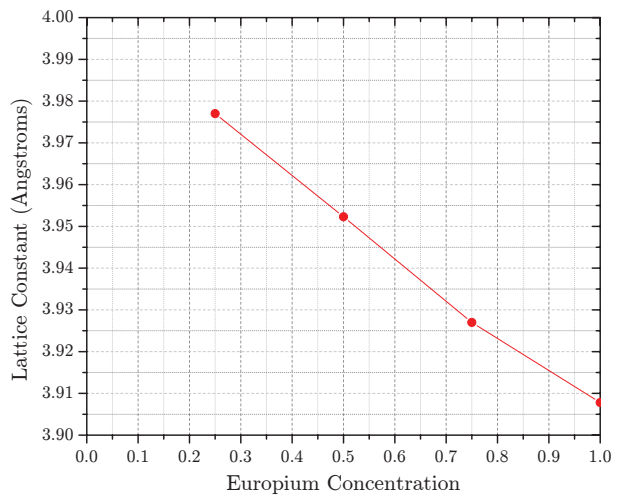}
    \caption{The lattice constant dependence on Eu concentration in Eu$_{x}$Ba$_{1-x}$TiO$_3$ ceramics at 300~K. The line is a guide for the eye}
    \label{fig:XRD3}
\end{figure}

In its paraelectric phase, (Eu,Ba)TiO$_3$ has a $Pm\bar{3}m$ perovskite structure with cubic symmetry. We have taken X-ray diffraction (XRD) spectra of all of the synthesized ceramics at room temperature (when they are all paraelectric), see Fig.~\ref{fig:XRD1} for Eu$_{0.5}$Ba$_{0.5}$TiO$_3$ spectrum. The spectrum of BaTiO$_3$, which is ferroelectric at 300~K is shown for comparison. The lattice constant trend with changing Eu concentration is shown in Fig.~\ref{fig:XRD3}.

\subsection{Insulating properties}

\begin{figure}[h!]
    \includegraphics[width=\columnwidth]{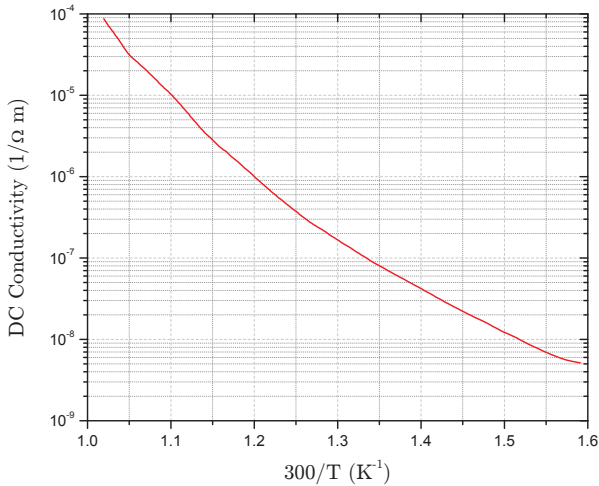}
    \caption{Eu$_{0.5}$Ba$_{0.5}$TiO$_3$ conductivity as the sample is cooled down.}
    \label{fig:cond}
\end{figure}

A material suitable for an EDM search has to be a good insulator. We measured sample resistance using an Agilent model 34410A digital multimeter as the sample was being cooled by flowing cold nitrogen vapor inside a Janis model 10CNDT dewar.
We found that EuTiO$_3$ and Eu$_{0.75}$Ba$_{0.25}$TiO$_3$ exhibited conductivity losses all the way to 4.2~K, and are therefore unsuitable for an EDM search. However for the Eu$_{0.5}$Ba$_{0.5}$TiO$_3$ and Eu$_{0.25}$Ba$_{0.75}$TiO$_3$ compositions, the conductivity losses were too small to detect already at 180~K, see Figure~\ref{fig:cond}. No measurements were done at lower temperature since the multimeter saturated, but the smallness of the conductivity losses at 77~K and 4~K can be inferred from the frequency-independence of the ferroelectric hysteresis loops.

\subsection{Ferroelectric properties and the effective electric field}

\begin{figure}[h!]
    \includegraphics[width=\columnwidth]{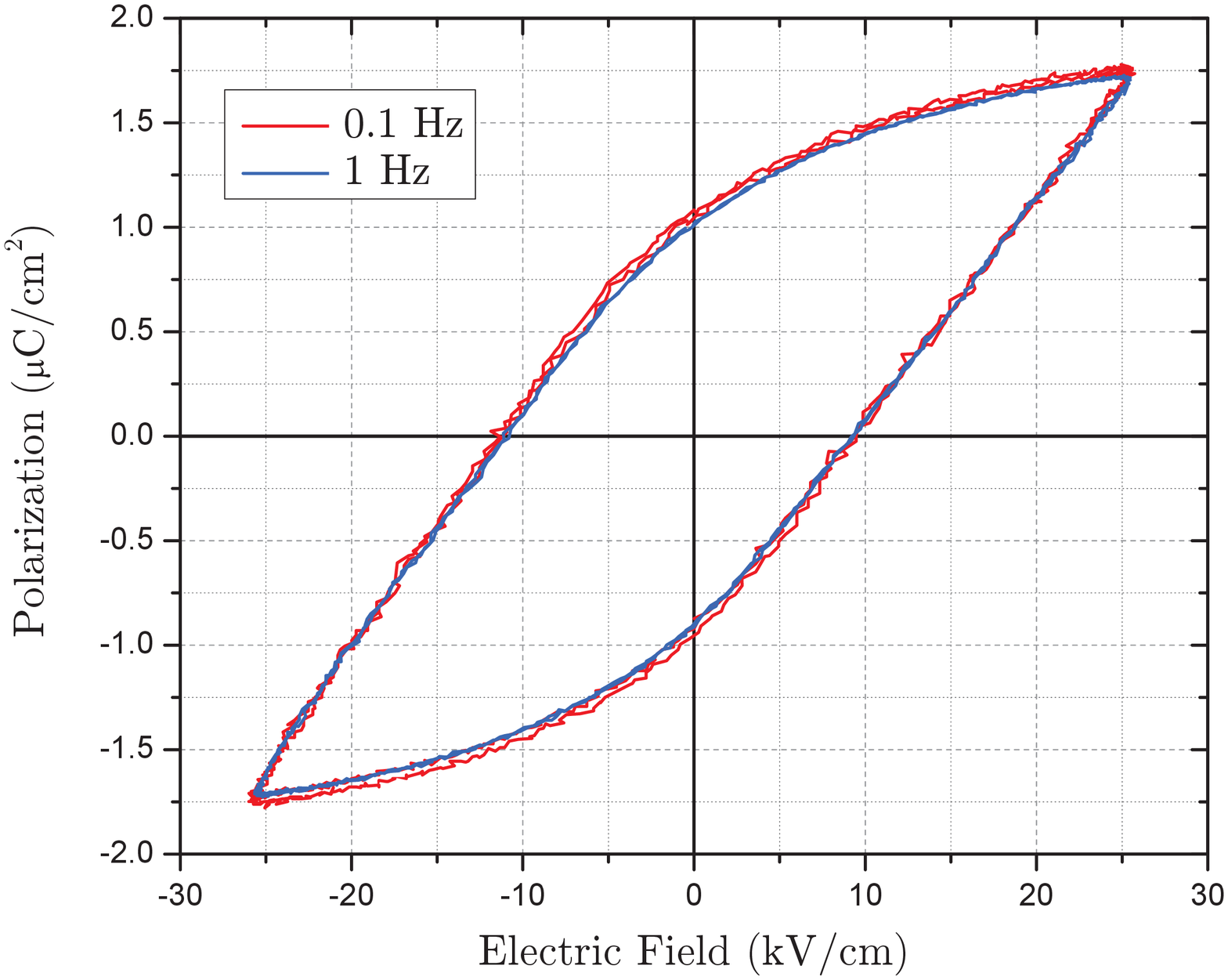}
    \caption{Eu$_{0.5}$Ba$_{0.5}$TiO$_3$ ferroelectric hysteresis loops at 4~K (immersed in liquid helium). Different curves correspond to different frequencies of applied electric field, they lie very nearly on top of each other.}
    \label{fig:hyst4}
\end{figure}

Although EuTiO$_3$ remains paraelectric down to absolute zero~\cite{Katsufuji2001}, addition of Ba into the lattice leads to the appearance of a ferroelectric phase, with the transition temperature increasing with greater Ba concentration. At this stage we have been unable to find the exact values of ferroelectric transition temperatures in the Eu$_x$Ba$_{1-x}$TiO$_3$ ceramics that we have prepared, but we have been able to verify that some of them are indeed ferroelectric at 4~K and 77~K, by measuring their ferroelectric hysteresis loops.
We were unable to make any ferroelectric hysteresis measurements with EuTiO$_3$ and Eu$_{0.75}$Ba$_{0.25}$TiO$_3$ compositions since they exhibited conductive AC losses. We have measured the ferroelectric properties of Eu$_{0.5}$Ba$_{0.5}$TiO$_3$ at 77~K and 4~K, the hysteresis loops measured at 4~K are shown in Figure~\ref{fig:hyst4}. Data were taken with a triangle-waveform applied at two different frequencies: 0.1~Hz and 1~Hz. The hysteresis loop is the same for the two frequencies, which implies that the conductivity losses are not detectable at this temperature.

The effective electric field $E^*$, giving rise to an energy shift in the presence of the electron EDM, can be estimated from the remanent polarization of
\begin{equation}
\label{eq:P0} P_0 \approx 1\text{ }\mu\text{C/cm}^2.
\end{equation}
We assume the following relationship between the ferroelectric displacements in the unit cell~\cite{Kwei1993,Kostja2009}:
\begin{equation}
\label{eq:x1} x_{Eu-O}=1.7x_{Ba-O}=x_{Ti-O}/2,
\end{equation}
where all the displacements are with respect to the center of the oxygen octahedra. Taking into account our ceramic density of 60\%, this gives the mean Eu-O displacement along the electric field:
\begin{equation}
\label{eq:x2} x_{Eu-O}\approx 0.01\AA.
\end{equation}
The EDM-induced energy shift due to this displacement is given by
~\cite{Mukhamedjanov2003}:
\begin{equation}
\label{eq:delta} \delta=-0.1\frac{x_{Eu-O}}{a_B}\frac{d_e}{ea_B}\times 27.2\text{ eV},
\end{equation}
where $a_B$ is the Bohr magneton, and $e$ is the magnitude of the electron's charge.
This can be converted to the effective electric field using $\delta=-d_eE^*$, resulting in
\begin{equation}
\label{eq:E*} E^*\approx 10\text{ MV/cm}.
\end{equation}

\subsection{Magnetic susceptibility and magnetization noise}

\begin{figure}[h!]
    \includegraphics[width=\columnwidth]{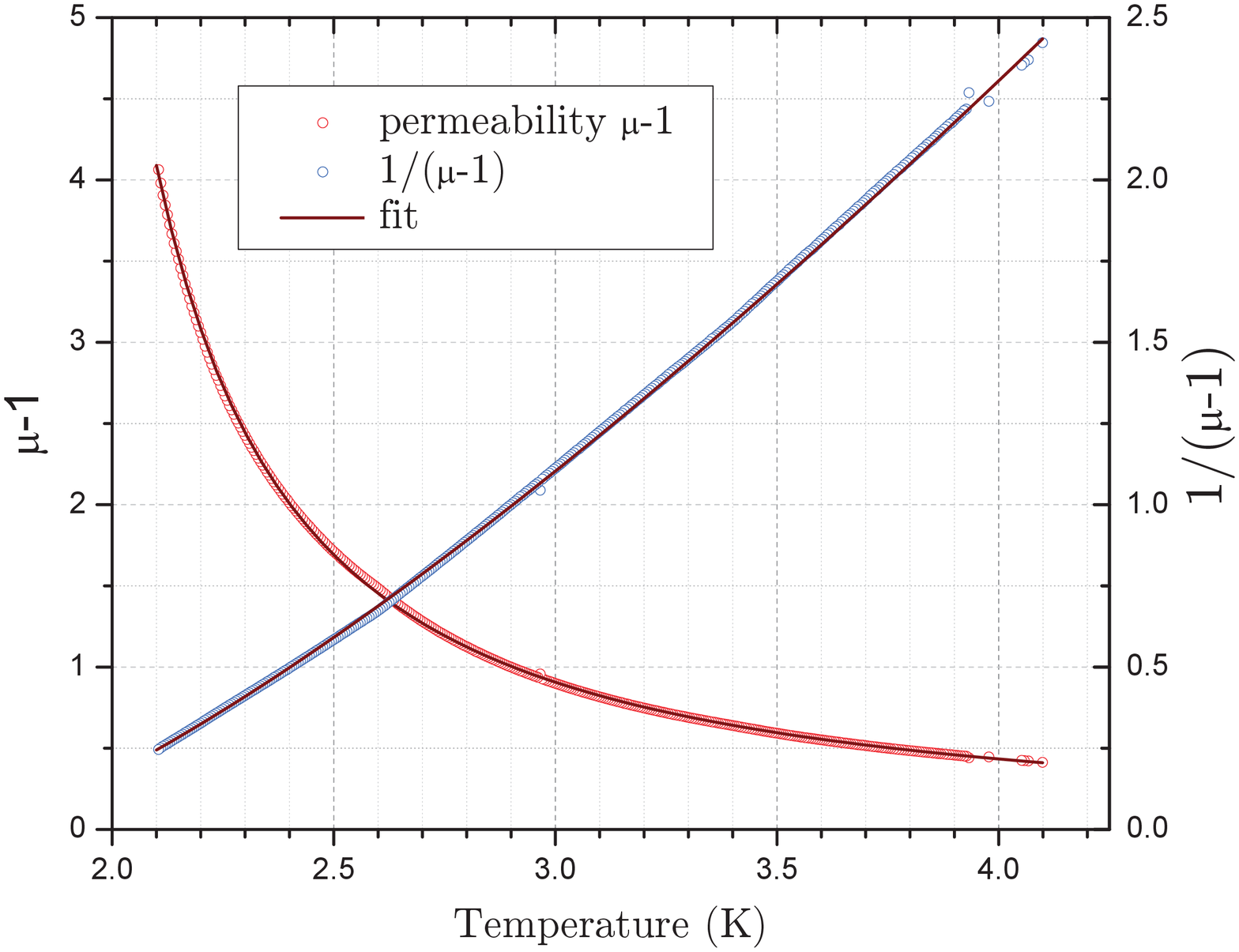}
    \caption{Eu$_{0.5}$Ba$_{0.5}$TiO$_3$ permeability as a function of temperature. The data is fit with the model in Eq.~(\ref{eq:mu1}).}
    \label{fig:perm}
\end{figure}

We measured the permeability $\mu-1$ of Eu$_{0.5}$Ba$_{0.5}$TiO$_3$ near 4~K by cutting a hole in one of the ceramic disk samples and wrapping, with copper wire, a 51-turn inductor primary winding, with the sample as the toroidal core. A one-turn niobium loop was wrapped on top of this toroid as the secondary winding and connected to a Quantum Design DC SQUID magnetometer. The setup was cooled inside a Janis model 10CNDT liquid helium dewar, and a vacuum pump was used to pump on the helium bath to reduce the temperature below 4.2~K. A known current was passed through the copper primary winding, and the response of the SQUID was detected. The permeability of the ceramic material is then calculated from the measured mutual inductance of the two loops. The results are shown in Figure~\ref{fig:perm}.


It is most likely that Eu$_{0.5}$Ba$_{0.5}$TiO$_3$ is a spin-glass, but more measurements need to be performed to ascertain the nature of the magnetic ordering in this material. We have fit the following function form to the permeability data:
\begin{equation}
\label{eq:mu1} \mu-1 = \frac{A}{(T-\theta)^{\gamma}},
\end{equation}
with the fit resulting in parameter values
\begin{equation}
\label{eq:mu2} A = 1.23\pm 0.01,\, \theta = (1.69\pm0.01)\text{ K},\, \gamma = 1.30\pm 0.01.
\end{equation}
It should be noted that the quoted errors are from the fit only, we estimate additional errors, due to flux leakage and SQUID mis-calibration, on the order of 10\%.

The presence of excessive magnetization noise in the magnetic material under study can degrade magnetic field sensitivity and weaken the achievable EDM limit~\cite{Sushkov2009, Eckel2009}. The setup described above was used to look for magnetization noise in Eu$_{0.5}$Ba$_{0.5}$TiO$_3$ at 4.2~K, but it was found that this noise is at or below the SQUID sensitivity. Therefore our magnetic field sensitivity of $\delta B \approx 3$~fT/$\rtHz$ is limited by the SQUID magnetometer noise.

%
%
%

\section{Conclusion}

We have synthesized Eu$_x$Ba$_{1-x}$TiO$_3$ ceramics and conducted a preliminary study of their properties. We conclude that Eu$_{0.5}$Ba$_{0.5}$TiO$_3$ is a suitable candidate for an electron EDM search, which has the potential to improve the current limit on the electron EDM by an order of magnitude.

\section*{Acknowledgements}
The authors would like to acknowledge useful discussions with Dmitry Budker, Louis Bouchard, David DeMille, Oleg Sushkov, Nicola Spaldin, Konstantin Rushchanskii, Marjana Lezaic, Stanislav Kamba, Dominic Ryan, Juergen Haase, Nataliya Georgieva, Mohana Yethiraj, Anna Mulders, and Chris Wiebe. The authors would like to acknowledge assistance from Zhenting Jiang, Jim Eckert, Vincent Bernardo, and the Gibbs machine shop team. This work was supported by Yale University.

\bibliography{References}

\end{document}